\documentclass[runningheads]{llncs}
\usepackage{amsmath}

\usepackage[T1]{fontenc}
\usepackage{hyperref}

\usepackage{graphicx}
\usepackage{color}

\usepackage{tikz}
\usetikzlibrary{arrows,automata,positioning}

\usepackage{amssymb}

\newcommand{\access}{\mathsf{access}}
\newcommand{\rank}{\mathsf{rank}}
\newcommand{\select}{\mathsf{select}}

\begin{document}
\title{Fast Pattern Matching with Epsilon Transitions}
\author{Nicola Cotumaccio \orcidID{0000-0002-1402-5298}}
\authorrunning{N. Cotumaccio}
\institute{Department of Computer Science, University of Helsinki, Finland
\email{nicola.cotumaccio@helsinki.fi}}
\maketitle              % typeset the header of the contribution
\begin{abstract}
In the String Matching in Labeled Graphs (SMLG) problem, we need to determine whether a pattern string appears on a given labeled graph or a given automaton. Under the Orthogonal Vectors hypothesis, the SMLG problem cannot be solved in subquadratic time [ICALP 2019]. In typical bioinformatics applications, pattern matching algorithms should be both fast and space-efficient, so we need to determine useful classes of graphs on which the SLMG problem can be solved efficiently.

In this paper, we improve on a recent result [STACS 2024] that shows how to solve the SMLG problem in linear time on the compressed representation of Wheeler generalized automata, a class of string-labeled automata that extend de Bruijn graphs. More precisely, we show how to remove the assumption that the automata contain no $ \epsilon $-transitions (namely, edges labeled with the empty string), while retaining the same time and space bounds. This is a significant improvement because $ \epsilon $-transitions add considerable expressive power (making it possible to jump to multiple states for free) and capture the complexity of regular expressions (through Thompson's construction for converting a regular expression into an equivalent automaton). We prove that, to enable $ \epsilon $-transitions, we only need to store two additional bitvectors that can be constructed in linear time.

\end{abstract}

\section{Introduction}

Pattern matching is one of the fundamental problems in computer science. In the 1970s, two elegant linear-time solutions were proposed. One solution is based on the Knuth-Morris-Pratt (KMP) algorithm \cite{knuth1977}; the other solution is based on suffix trees \cite{weiner1973}. Twenty years later, several authors extensively studied the \emph{String Matching in Labeled Graphs (SMLG)} problem, which is the natural generalization of pattern matching to a graph setting \cite{manber1992,tasuya1993,park1995,amir2000,rautiainen2017,navarro2000}. Loosely speaking, the SMLG problem can be defined as follows: given a directed graph whose nodes (or edges) are labeled with strings and given a pattern string, decide whether the pattern can be read by following a path on the graph and concatenating the labels encountered along the way. Since every text can be seen as a graph consisting of a single path, the SMLG problem is a natural extension of the classical pattern matching problem. Unfortunately, the elegant and general solutions based on the KMP algorithm and suffix trees do not admit a simple generalization to the graph setting. Amir et al. \cite{amir2000} showed that the (exact) SMLG problem on arbitrary graphs can be solved in $ O(\mathfrak{e} + me) $ time, where $ e $ is the number of edges in the graph, $ m $ is the length of the pattern, and $ \mathfrak{e} $ is the total length of all labels in the graph. In the (approximate) variant where one allows errors in the graph, the problem becomes NP-hard \cite{amir2000}, so generally errors are only allowed in the pattern. Recently, some conditional lower bounds showed that the SMLG problem is inherently more complex than the classical pattern matching problem. Equi et al. \cite{equiicalp2019,equi2021} showed that, given an arbitrary graph, the SMLG cannot be solved in $ O(me^{1 - \epsilon}) $ or $ O(m^{1 - \epsilon}e) $ time (for every $ \epsilon > 0 $), unless the Orthogonal Vectors hypothesis is not true.

Due to the sheer increase in the amount of data (and in particular of genomic data), in applications (and especially in bioinformatics) we need algorithms that are both fast and space-efficient. A polynomial algorithm for the SMLG problem running in $ O(me) $ time cannot be satisfactory, so the SMLG problem has been restricted to classes of graphs on which it can be solved more efficiently. For example, the SMLG problem can be solved in linear time on \emph{Elastic Founder graphs} under some non-restrictive assumptions \cite{equiefg2023,rizzo2022}. Elastic Founder graphs are used to represent multiple sequence alignments (MSA), a central model of biological evolution.

We mentioned that we also strive for space-efficient algorithms. The pattern matching problem \emph{on texts} has been revolutionized by the invention of the Burrows-Wheeler Transform \cite{burrows1994} and the FM-index \cite{ferraginajacm2005}, which allow solving pattern matching queries efficiently on \emph{compressed} text, thus establishing a new paradigm in bioinformatics. Over the last decade, these ideas were extended to labeled graphs and automata. On the popular class of \emph{Wheeler automata}, the SMLG problem can be solved in linear time while only storing a compressed representation of the Wheeler automaton \cite{gagie2017,alanko2020}. Note that a labeled graph and an automaton are essentially the same object: an automaton is simply a labeled graph with an initial state and a set of final states. However, we will generally speak of automata rather than graphs because the results in this area often have relevant applications in formal language theory, and because some (algorithmic) properties of labeled graphs are tied to determinism. For example, the powerset construction applied to a Wheeler non-deterministic automaton leads to a linear blow-up in the number of states of the equivalent deterministic automaton, and the equivalent deterministic automaton is Wheeler \cite{alanko2020}; on arbitrary non-deterministic automata, the blow-up can be exponential. A special case of Wheeler automata are de Bruijn graphs \cite{bowe2012}, which are used to perform Eulerian sequence assembly \cite{idury1995,pavel2001,bankevich2012}.

\begin{figure}[h]
     \centering
        \scalebox{1}{
        \begin{tikzpicture}[->,>=stealth', semithick, auto, scale=1]
%\tikzset{every state/.style={minimum size=0pt}}
\node[state, initial, initial where=below] (1)    at (0,0)	{$ 1 $};
\node[state] (7)    at (-5, 2)	{$ 7 $};
\node[state] (9)    at (-3, 2)	{$ 9 $};
\node[state] (10)    at (-1, 2)	{$ 10 $};
\node[state] (2)    at (1, 2)	{$ 2 $};
\node[state] (8)    at (3, 2)	{$ 8 $};
\node[state] (6)    at (5, 2)	{$ 6 $};
\node[state, accepting] (4)    at (-4, 4)	{$ 4 $};
\node[state] (5)    at (0, 4)	{$ 5 $};
\node[state, accepting] (3)    at (4, 4)	{$ 3 $};

\draw (1) edge [bend left] node [] {$ b $} (7);
\draw (1) edge [] node [] {$ bb $} (9);
\draw (1) edge [] node [] {$ c $} (10);
\draw (1) edge [] node [] {$ a $} (2);
\draw (1) edge [] node [] {$ bb $} (8);
\draw (1) edge [bend right] node [] {$ b $} (6);
\draw (7) edge [] node [] {$ ba $} (4);
\draw (9) edge [] node [] {$ a $} (4);
\draw (10) edge [] node [] {$ ca $} (5);
\draw (2) edge [] node [] {$ ca $} (5);
\draw (8) edge [] node [] {$ a $} (3);
\draw (6) edge [] node [] {$ ba $} (3);
\draw (5) edge [] node [] {$ \epsilon $} (4);
\draw (5) edge [] node [] {$ \epsilon $} (3);
\end{tikzpicture}
}
\caption{A Wheeler $ 2 $-GNFA. The states are numbered following the Wheeler order.}
\label{fig:example}
\end{figure}
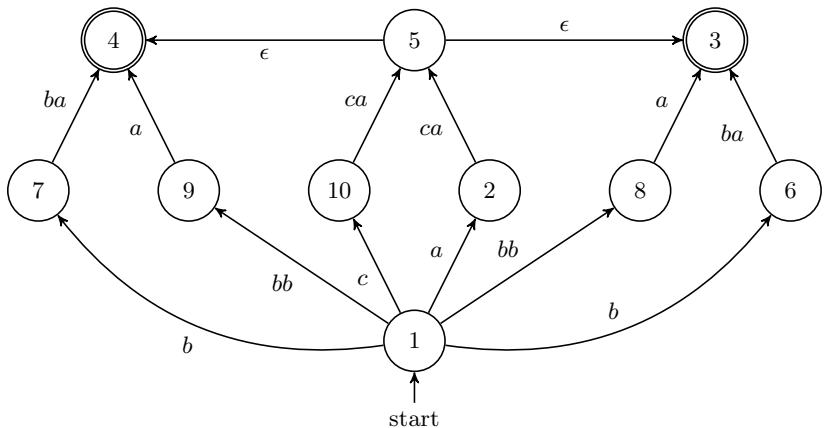  

\subsubsection{Our results} Recently, the notion of Wheelerness was extended to \emph{generalized automata} (see Figure \ref{fig:example}), namely, automata where edges can be labeled not only with characters, but also strings \cite{cotumaccio2024stacs}. This is a natural step, because in data compression and bioinformatics many common data structures are based on string-labeled graphs, such as Patricia trees, suffix trees and pangenomes (as well as Elastic Founder graphs) \cite{navarro2016,baaijens2022,makinen2023}. We say that a generalized non-deterministic finite automaton (GNFA) is an \emph{$ r $-GNFA} if all edge labels have length at most $ r $ (so a GNFA is a conventional NFA if and only if it is an $ 1 $-GNFA). Let $ m $, $ e $ and $ \mathfrak{e} $ as above, and let $ \sigma = |\Sigma| $. Consider the following result.

\begin{theorem}\label{theor:fmindex}
    Let $ \mathcal{A} $ be a Wheeler $ r $-GNFA, and assume that $ r = O(1) $. Then, we can encode $ \mathcal{A} $ by using $ \mathfrak{e} \log \sigma (1 + o(1)) + O(e) $ bits so that later on, given a pattern $ \alpha \in \Sigma^* $ of length $ m $, we can solve the SMLG problem on $ \mathcal{A} $ in $ O(m \log \log \sigma) $ time. Within the same time bound, we can also decide whether $ \alpha $ is accepted by $ \mathcal{A} $.
\end{theorem}

Theorem \ref{theor:fmindex} states that on Wheeler GNFAs we can solve pattern matching queries in linear time (for constant alphabets), thus overcoming the lower bound of Equi et al. while only storing a compressed representation of $ \mathcal{A} $. The case $ r = 1 $ corresponds to conventional Wheeler automata \cite{gagie2017}. This theorem was proved in  \cite{cotumaccio2024stacs} under the crucial assumption that $ \mathcal{A} $ contains no $ \epsilon $-transitions (that is, edges labeled with the empty string $ \epsilon $): a typical pattern matching algorithm runs in time proportional to the pattern length $ m $, which is not suitable in the context of $ \epsilon $-transitions because the empty string has length equal to zero.

In this paper, we show that, surprisingly, we can achieve the time and space bounds in Theorem \ref{theor:fmindex} \emph{even in the presence of $ \epsilon $-transitions}, without imposing any restrictions on the automaton $\mathcal{A} $. More precisely, we will show that we only need to store two additional bitvectors. Intuitively, these bitvectors capture the range of states that we can reach by following $ \epsilon $-transitions. We also show that these bitvectors can be built in linear time (Theorem \ref{theor:construction}). In addition to proving the result in \cite{cotumaccio2024stacs} in full generality, supporting $ \epsilon $-transitions provides some specific advantages.

\begin{itemize}
    \item $ \epsilon $-transitions add considerable expressive power to the set of all strings captured by $ \mathcal{A} $ because they enable jumping from one state to another for free (one can potentially follow multiple $ \epsilon $-transitions at once for free).
    \item $ \epsilon $-transitions already have a crucial role in the context of conventional automata. Wheeler automata and related automata capture the complexity of regular expression under standard operators such as union and complementation \cite{cotumacciojacm}, and Thompson's algorithm, the standard construction for converting a regular expression into an automaton, heavily relies on $ \epsilon $-transitions \cite{hopcroft2006}. 
\end{itemize}

The paper is organized as follows. Section \ref{sec:preliminaries} introduces our notation and our first definitions. Section \ref{sec:howtosolve} outlines the proof in \cite{cotumaccio2024stacs} for GNFAs without $ \epsilon $-transitions. In Section \ref{sec:patternepsilon}, we prove Theorem \ref{theor:fmindex}. In Section \ref{sec:construction}, we prove Theorem \ref{theor:construction} (linear-time construction of the two bitvectors). Section \ref{sec:conclusions} presents our conclusions.

Due to space constraints, some proofs can be found in the appendix.

\section{Preliminaries}\label{sec:preliminaries}

\subsubsection{First definitions}

Let $ V $ be a set. We say that a (binary) relation $ \le $ on $ V $ is a \emph{partial} order if $ \le $ is reflexive, antisymmetric and transitive. For every $ u, v \in V $, we write $ u < v $ if $ (u \le v) \land (u \not = v) $. A partial order $ \le $ is a \emph{total} order if for every $ u, v \in V $ we have $ (u \le v) \lor (v \le u) $. We say that $ U \subseteq V $ is \emph{$ \le $-convex} if, for every $ u, v, z \in V $, from $ u < v < z $ and $ u, z \in U $ we obtain $ v \in U $.

Let $ \Sigma $ be a finite alphabet. We denote by $ \Sigma^* $ the set of all finite strings on $ \Sigma $, where $ \epsilon \in \Sigma^* $ is the empty string. For $ i \ge 0 $, let $ \Sigma^i \subseteq \Sigma^* $ be the set of all strings of length $ i $ on the alphabet $ \Sigma $. If $ \alpha, \beta \in \Sigma^* $, we write $ \alpha \dashv \beta $ if and only if $ \alpha $ is a suffix of $ \beta $ (equivalently, if and only if there exists $ \beta' \in \Sigma^* $ such that $ \beta = \beta' \alpha $). If $ \alpha \in \Sigma^* $ and $ 0 \le k \le |\alpha| $, let $ p(\alpha, k) $ and $ s(\alpha, k) $ be the prefix and the suffix of $ \alpha $ of length $ k $, respectively. We assume that there exists a fixed total order $ \preceq $ on the alphabet $ \Sigma $ (in our examples, we always assume $ a \prec b \prec c \prec \dots $), and $ \preceq $ is extended \emph{co-lexicographically} to $ \Sigma^* $ (that is, for every $ \alpha, \beta \in \Sigma^* $ we have $ \alpha \prec \beta $ if and only if the reverse string $ \alpha^R $ is lexicographically smaller than the reverse string $ \beta^R $).

A \emph{generalized non-deterministic finite automaton (GNFA)} is a 4-tuple $ \mathcal{A} = (Q, E, s, F) $, where $ Q $ is a finite set of states, $ E \subseteq Q \times Q \times \Sigma^* $ is a finite set of string-labeled edges, $ s \in Q $ is the initial state and $ F \subseteq Q $ is a set of final states (see Figure \ref{fig:example}). Moreover, we assume that each $ u \in Q $ is reachable from the initial state and is co-reachable, that is, it is either final or allows reaching a final state. This is a standard assumption because all states that do not satisfy this requirement can be removed without changing the language recognized by the automaton. We say that a GNFA $ \mathcal{A} = (Q, E, s, F)  $ is a  \emph{GNFA without $ \epsilon $-transitions} if for every $ (u, v, \rho) \in E $ we have $ \rho \not = \epsilon $. A \emph{generalized deterministic finite automaton (GDFA)} is a GNFA without $ \epsilon $-transitions $ \mathcal{A} = (Q, E, s, F)  $ such that for every pair of \emph{distinct} edges $ (u, v, \rho), (u', v', \rho') \in E $, if $ u = u' $, then $ \rho $ is not a prefix of $ \rho' $ and $ \rho' $ is a not a prefix of $ \rho $ (in particular, $ \rho \not = \rho' $).

Let $ \mathcal{A} = (Q, E, s, F) $ be a GNFA. For $ r \ge 0 $, we say that $ \mathcal{A} $ is an $ \emph{r-GNFA} $ if each edge label has length at most $ r $ (namely, $ |\rho| \le r $ for every $ (u, v, \rho) \in E $). For $ U \subseteq Q $ and $ \rho \in \Sigma^* $, let $\mathtt{out}(U,\rho)$ be the number of edges labeled with $ \rho $ that leave states in $ U $, and let $\mathtt{in}(U,\rho)$ be the number of edges labeled with $ \rho $ that enter states in $ U $. For every $ u \in Q $ let $ \lambda (u) $ be the set of all strings in $ \Sigma^* $ labeling an edge reaching $ u $.
%we denote by $ \min_{\lambda (u)} $ and $ \max_{\lambda (u)} $ the (co-lexicographically) smallest and largest strings in $ \lambda (u) $, respectively.
Moreover, for every $ u \in Q $, let $ I_u $ be the set of all strings that can be read from the initial state to $ u $ by concatenating edge labels. In other words, for every $ \alpha \in \Sigma^* $ we have $ \alpha \in I_u $ if and only, for some $ t \ge 1 $, there exist $ u_1, \dots, u_t \in Q $ and $ \alpha_1, \dots, \alpha_{t - 1} \in \Sigma^* $ such that (i) $ u_1 = s $ and $ u_t = u $, (ii) $ (u_i, u_{i + 1}, \alpha_i) \in E $ for every $ 1 \le i \le t - 1 $, and (iii) $ \alpha = \alpha_1 \alpha_2 \dots \alpha_{t - 1} $ (in particular, $ \epsilon \in I_s $). As observed in \cite{cotumaccio2024stacs}, if $ \mathcal{A} = (Q, E, s, F) $ is a GDFA and $ u, v \in Q $, with $ u \not = v $, then $ I_u \cap I_v = \emptyset $.

Let $ \mathcal{A} $ be a GNFA. The \emph{String Matching in Labeled Graphs (SMLG)} problem for GNFAs is defined as follows: build a data structure that encodes $ \mathcal{A} $ such that, given a string $ \alpha $, we can efficiently compute the set of all states reached by a path suffixed by $ \alpha $.

Our main result (Theorem \ref{theor:fmindex}) holds for polynomial alphabets (namely, $ \sigma \le e^{O(1)} $) in the word RAM model with words of size $ w \in  \Theta(\log N) $ bits, where $ N $ is the input size.

\subsubsection{Wheeler GNFAs}

Let $ \mathcal{A} = (Q, E, s, F) $ be a GNFA. Let $ \preceq_\mathcal{A} $ be the relation on $ Q $ such that, for every $ u, v \in Q $, we have $ u \preceq_\mathcal{A} v $ if and only if $ (\forall \alpha \in I_u)(\forall \beta \in I_v)((\{\alpha, \beta \} \not \subseteq I_u \cap I_v )\to (\alpha \prec \beta)) $. As observed in \cite{cotumaccio2024stacs}, in general $ \preceq_\mathcal{A} $ is only a \emph{preorder}, that is, it is a reflexive and transitive relation, but it need not be antisymmetric. If $ \mathcal{A} $ is a GDFA, we know that $ u \not = v $ implies $ I_u \cap I_v = \emptyset $, so $ \preceq_\mathcal{A} $ is a partial order, and it is the reflexive relation such that, for every distinct $ u, v \in Q $ we have $ u \prec_\mathcal{A} v $ if and only if $ (\forall \alpha \in I_u)(\forall \beta \in I_v)(\alpha \prec \beta) $.

Let $ \mathcal{A} = (Q, E, s, F) $ be a GNFA. We say that $ \mathcal{A} $ is \emph{Wheeler} \cite{cotumaccio2024stacs} if there exists a total order $ \le $ on $ Q $ such that:
    \begin{itemize}
        \item \emph{(Axiom 1)} For every $ u, v \in Q $, if $ u \le v $, then $ u \preceq_\mathcal{A} v $.
        \item \emph{(Axiom 2)} $ s $ comes first in the total order $ \le $.
        \item \emph{(Axiom 3)} For every $ (u', u, \rho), (v', v, \rho') \in E $, if $ u < v $ and $ \rho' $ is not a strict suffix of $ \rho $, then $ \rho \preceq \rho' $.
        \item \emph{(Axiom 4)} For every $ (u', u, \rho), (v', v, \rho) \in E $, if $ u < v $, then $ u' \le v' $.
    \end{itemize}
We say that $ \le $ is a \emph{Wheeler order} on $ \mathcal{A} $. (see Figure \ref{fig:example} for an example).

Notice that, if $ \mathcal{A} $ is a GNFA without $ \epsilon $-transitions, then Axiom 2 follows from Axiom 1 because $ \epsilon \in I_s $ and for every $ u \in Q \setminus \{s \} $ we have $ \epsilon \not \in I_u $. Moreover, if $ \mathcal{A} $ is a GDFA, then Axioms 2, 3, 4 follow from Axiom 1 (see \cite{cotumaccio2024stacs}). Lastly, if $ \mathcal{A} $ is a conventional NFA, then Axiom 1 follows from Axiom 2, 3, 4 (see \cite{cotumaccio2024stacs}) and we retrieve the usual \emph{local} definition of Wheeler NFA \cite[Corollary 3.1]{alanko2020}: a (conventional) NFA is Wheeler if there exists a total order $ \le $ on $ Q $ such that (i) $ s $ comes first in the total order $ \le$, (ii) for every $ (u', u, a), (v', v, b) \in E $, if $ u < v $, then $ a \preceq b $ and (iii) for every $ (u', u, a), (v', v, a) \in E $, if $ u < v $, then $ u' \le v' $.

If $ \mathcal{A} = (Q, E, s, F) $ is a GNFA and $ \le $ is a Wheeler order on $ \mathcal{A} $, we write $ Q = \{Q[1], Q[2], \dots, Q[|Q|] \} $, where $ Q[1] < Q[2] < \dots < Q[|Q|] $. If $ 1 \le i \le j \le |Q| $, let $ Q[i, j] = \{Q[i], Q[i + 1], \dots, Q[j - 1], Q[j] \} $, and if $ i > j $, let $ Q[i, j] = \emptyset $.

\section{How to Solve Pattern Matching Queries}\label{sec:howtosolve}

In \cite{cotumaccio2024stacs}, Theorem \ref{theor:fmindex} was proved under the assumption that $ \mathcal{A} $ is a Wheeler GNFA without $ \epsilon $-transitions. This section briefly summarizes the proof in \cite{cotumaccio2024stacs} to highlight the technical contributions required to remove the assumption that $ \mathcal{A} $ does not have $ \epsilon $-transitions. We start with a definition.

\begin{definition}
Let $ \mathcal{A} = (Q, E, s, F) $ be a Wheeler GNFA, and let $ \alpha \in \Sigma^* $. Define:
\begin{itemize}
    \item $ G^\prec(\alpha) = \{  u \in Q \;|\; (\forall \beta\in I_{u})(\beta \prec \alpha)\} $;
    \item $ G_\dashv(\alpha) = \{  u \in Q \;|\; (\exists \beta\in I_{u})( \alpha \dashv \beta)\} $;
    \item $ G^\prec_\dashv(\alpha) = G^\prec(\alpha) \cup G_\dashv(\alpha) =\{  u \in Q\ \;|\; (\forall \beta\in I_{u})(\beta \prec \alpha) \vee (\exists \beta\in I_{u})( \alpha \dashv  \beta)\} $.
\end{itemize}
\end{definition}

The set $ G_\dashv(\alpha) $ is the set of states that the SMLG problem must return on input $ \alpha $. The following lemma shows, among other properties, that $ G_\dashv(\alpha) = Q[|G^\prec(\alpha)| + 1, |G^\prec_\dashv(\alpha)|] $. This result was proved in \cite{cotumaccio2024stacs} for Wheeler GNFAs without $ \epsilon $-transitions, but it can be extended readily to arbitrary Wheeler GNFAs.

\begin{lemma}\label{lem:intervals}
Let $ \mathcal{A} = (Q, E, s, F) $ be a Wheeler GNFA, let $ \le $ be a Wheeler order on $ \mathcal{A} $, and let $ \alpha \in \Sigma^* $. Then:
\begin{enumerate}
    \item $ G^\prec(\alpha) \cap G_\dashv(\alpha) = \emptyset $.
    \item $ G_\dashv(\alpha) $ is $ \le $-convex.
    \item If $ u, v \in Q $ are such that $ u < v $ and $ v \in G^\prec(\alpha) $, then $ u \in G^\prec(\alpha) $. In other words, $ G^\prec(\alpha) = Q[1, |G^\prec(\alpha)|] $.
    \item If $ u, v \in Q $ are such that $ u < v $ and $ v \in G^\prec_\dashv(\alpha) $, then $ u \in G^\prec_\dashv(\alpha) $. In other words, $ G^\prec_\dashv(\alpha) = Q[1, |G^\prec_\dashv(\alpha)|] $.
    \item $ G_\dashv(\alpha) = Q[|G^\prec(\alpha)| + 1, |G^\prec_\dashv(\alpha)|] $.
\end{enumerate}
\end{lemma}

Since $ G_\dashv(\alpha) = Q[|G^\prec(\alpha)| + 1, |G^\prec_\dashv(\alpha)|] $, to compute $ G_\dashv(\alpha) $ we only need to compute $ |G^\prec(\alpha)| $ and $ |G^\prec_\dashv(\alpha)| $. To this end, we recursively compute $ |G^\prec(p(\alpha, k))| $ and $ |G^\prec_\dashv(p(\alpha, k))| $ for every $ 0 \le k \le |\alpha| $, and the conclusion follows for $ k = |\alpha| $. Note that, for $ k = 0 $ we have $ p(\alpha, k) = \epsilon $, and we have $ |G^\prec(\epsilon)| = 0 $ and $ |G^\prec_\dashv(\epsilon)| = |Q| $ because $ \epsilon $ is the smallest string in $ \Sigma^* $ and $ \epsilon $ is a suffix of every string in $ \Sigma^* $.

We are only left with showing how to compute $ |G^\prec(\alpha)| $ and $ |G^\prec_\dashv(\alpha)| $, assuming that we have already computed $ |G^\prec(p(\alpha, k))| $ and $ |G^\prec_\dashv(p(\alpha, k))| $ for every $ 0 \le k < |\alpha| $.

Let us describe the approach in \cite{cotumaccio2024stacs} for Wheeler GNFAs without $ \epsilon $-transitions. First, let us show how to compute $ |G^\prec(\alpha)| $. Fix any $ Q[j] \in Q $. We know that $ Q[j] \in G^\prec(\alpha) $ if and only if for every $ \beta \in I_{Q[j]} $ we have $ \beta \prec \alpha $. As a consequence, if $ Q[j] \in G^\prec(\alpha) $, then for every $ \rho \in \lambda(Q[j]) $ we must have $ \rho \prec \alpha $. Moreover, if $ \rho $ is a suffix of $ \alpha $ --- say $ \rho = s(\alpha, k) $ --- then for every $ u' \in Q $ such that $ (u', Q[j], \rho) \in E $ we must have $ u' \in G^\prec(p(\alpha, |\alpha| - k)) $. Crucially, we have $ k \ge 1 $ (and not $ k \ge 0 $) because we are considering a GNFA without $ \epsilon $-transitions, so we have already computed $ |G^\prec(p(\alpha, |\alpha| - k))| $. Axiom 4 ensures that all edges labeled $ s(\alpha, k) $ ``respect'' the Wheeler order $ \le $ so, if $ f_k = \mathtt{out}(Q[1, |G^\prec(p(\alpha, |\alpha| - k))|], s(\alpha, k)) $, we have $ u' \in G^\prec(p(\alpha, |\alpha| - k)) $ for every $ u' \in Q $ such that $ (u', Q[j], s(\alpha, k)) \in E $ if and only if $ \mathtt{in}(Q[1, j], s(\alpha, k)) \leq f_k $. Intuitively, we obtain the following lemma (see \cite{cotumaccio2024stacs}).

\begin{lemma}\label{lem:lessold}
    Let $ \mathcal{A} = (Q, E, s, F) $ be a Wheeler $ r $-GNFA without $ \epsilon $-transitions, let $ \le $ be a Wheeler order on $ \mathcal{A} $, and let $ \alpha \in \Sigma^*$, with $ \alpha \not = \epsilon $. For $ 0 < k < \min\{r + 1,|\alpha|\} $, let $ f_k = \mathtt{out}(Q[1, |G^\prec(p(\alpha, |\alpha| - k))|], s(\alpha, k)) $. Then, $ |G^\prec(\alpha)| $ is the largest integer $ 0 \le j \le |Q| $ such that:
    \begin{itemize}
        \item $ \mathtt{in}(Q[1, j], s(\alpha, k)) \leq f_k $, for every $ 0 < k < \min\{r + 1,|\alpha|\} $;
        \item $ \rho \prec \alpha $ for every $ \rho \in \Sigma^k \cap \lambda (Q[h]) $, for every $ 1 \le h \le j $ and for every $ 0 < k < r + 1 $.
    \end{itemize}
\end{lemma}

Now, let us show how to compute $ |G^\prec_\dashv(\alpha)| $. Let $ G^*(\alpha) $ be the set of all states reached by an edge labeled with a string suffixed by $ \alpha $. Formally:
\begin{equation*}
        G^* (\alpha) = \{u \in Q \;|\; (\exists \rho \in \lambda (u))(\alpha \dashv \rho) \}.
\end{equation*}

Notice that $ G^* (\alpha) \subseteq G_\dashv (\alpha) $. Indeed, pick $ u \in G^* (\alpha) $. Then, there exists $ \rho \in \lambda (u) $ such that $ \alpha \dashv \rho $. In particular, there exists $ u' \in Q $ such that $ (u', u, \rho) \in E $. Pick any $ \beta \in I_{u'} $. Then, $ \beta \rho \in I_u $. From $ \alpha \dashv \rho $ we conclude $ \alpha \dashv \beta \rho $, so $ u \in G_\dashv (\alpha) $.

Consider any $ Q[j] \in Q $. Then, we have $ Q[j] \in G^\prec_\dashv(\alpha) $ if and only if $ Q[j] \in G^\prec(\alpha) $ or $ Q[j] \in G_\dashv(\alpha) $. Now, we have $ Q[j] \in G^\prec(\alpha) $ if and only if $ j \le |G^\prec(\alpha)| $, and we have already computed $ |G^\prec(\alpha)| $. We consider a GNFA without $ \epsilon $-transitions, so we have $ Q[j] \in G_\dashv(\alpha) $ if and only if either $ Q[j] \in G^*(\alpha) $, or there exists $ 1 \le k \le r $ and $ u' \in G_\dashv(p(\alpha, |\alpha| - k)) $ such that $ (u', Q[j], s(\alpha, k)) \in E $. Axiom 4 ensures that we can compute $ |G^\prec_\dashv(\alpha)| $ as follows (see \cite{cotumaccio2024stacs}).

\begin{lemma}\label{lem:lessequalold}
Let $ \mathcal{A} = (Q, E, s, F) $ be a Wheeler $ r $-GNFA without $ \epsilon $-transitions, let $ \le $ be a Wheeler order on $ \mathcal{A} $, and let $\alpha \in \Sigma^*$, with $ \alpha \not = \epsilon $. For $ 0 < k < \min\{r + 1,|\alpha|\} $, let $ f_k = \mathtt{out}(Q[1, |G^\prec(p(\alpha, |\alpha| - k))|], s(\alpha, k)) $ and let $ g_k = \mathtt{out}(Q[1, |G^\prec_\dashv(p(\alpha, |\alpha| - k))|], s(\alpha, k)) $. Then, $ g_k \ge f_k $ for every $ 0 < k < \min\{r + 1,|\alpha|\} $. Moreover, $ |G^\prec_\dashv(\alpha)| $ is the maximum among:
\begin{itemize}
    \item $ |G^\prec(\alpha)| $;
    \item the largest integer $ 0 \le i \le |Q| $ such that, if $ i \ge 1 $, then $ Q[i] \in G^* (\alpha) $;
    \item the smallest integer $ 0 \le j \le |Q| $ such that, for every $ 0 < k < \min\{r + 1,|\alpha|\} $ for which $ g_k > f_k $, we have $\mathtt{in}(Q[1, j], s(\alpha, k)) \ge g_k $.
\end{itemize}
\end{lemma}

For $ r = O(1) $ there exists a compressed data structure of $ \mathfrak{e} \log \sigma (1 + o(1)) + O(e) $ bits such that we can determine each condition in Lemma \ref{lem:lessold} and Lemma \ref{lem:lessequalold} in $ O(\log \log \sigma) $ time (for example, for a given $ 0 < k < \min\{r + 1,|\alpha|\} $, we can compute the largest integer $ 0 \le j \le |Q| $ such that $ \mathtt{in}(Q[1, j], s(\alpha, k)) \leq f_k $ in $ O(\log \log \sigma) $ time) \cite{cotumaccio2024stacs}. We conclude that, assuming that we have already computed $ |G^\prec(p(\alpha, k))| $ and $ |G^\prec_\dashv(p(\alpha, k))| $ for every $ 0 \le k < |\alpha| $, we can compute $ |G^\prec(\alpha)| $ and $ |G^\prec_\dashv(\alpha)| $ in $ O(\log \log \sigma) $ time. This yields the time bound $ O(m \log \log \sigma) $ in Theorem \ref{theor:fmindex} (where $ m = |\alpha| $) for Wheeler GNFAs without $ \epsilon $-transitions.

\section{Pattern Matching with $ \epsilon $-transitions}\label{sec:patternepsilon}

Let us show how to extend the results of Section \ref{sec:howtosolve} to arbitrary Wheeler GNFAs (possibly with $ \epsilon $-transitions). First, let us show that Lemma \ref{lem:lessold} is not true if $ \epsilon $-transitions are allowed. In Figure \ref{fig:example}, for $ \alpha = cba $, the largest integer $ 0 \le j \le |Q| $ such that the two conditions of Lemma \ref{lem:lessold} are true is $ 4 $ (note that $ |G^\prec(cb)| = 9 $, $ f_1 = 3 $, $ |G^\prec(c)| = 9 $, $ f_2 = 2 $), but $ |G^\prec(\alpha)| = 2 \not = 4 $. This is due to the $ \epsilon $-transitions $ (5, 3, \epsilon) $ and $ (5, 4, \epsilon) $ and intuitively, to obtain a correct variant of Lemma \ref{lem:lessold}, we need to remember the range of states that we can reach by following $ \epsilon $-transitions in a backward fashion. 

Let $ \mathcal{A} = (Q, E, s, F) $ be a GNFA. Given $ u, v \in Q $, we say that there exists an \emph{$ \epsilon $-walk} from $ u $ to $ v $ if, for some $ t \ge 1 $ there exist $ z_1, z_2, \dots, z_t \in Q $ such that (i) $ z_1 = u $ and $ z_t = v $ and (ii) $ (z_i, z_{i + 1}, \epsilon) \in E $ for every $ 1 \le i \le t - 1 $ (in particular, for every $ u \in Q $ there exists an $ \epsilon $-walk from $ u $ to $ u $).

We can now define $ A_{\max} $ and $ A_{\min} $, the two crucial arrays to obtain our result. 

\begin{definition}
    Let $ \mathcal{A} = (Q, E, s, F) $ be a Wheeler GNFA, and let $ \le $ be a Wheeler order on $ \mathcal{A} $. Let $ A_{\max}[1, |Q|] $ ($ A_{\min}[1, |Q|] $, respectively) be the array such that, for every $ 1 \le i \le |Q| $, we have that $ A_{max}[i] $ ($ A_{\min}[i] $, respectively) is the largest (smallest, respectively) integer $ 1 \le j \le |Q| $ for which there exists an $ \epsilon $-walk from $ Q[j] $ to $ Q[i] $.
\end{definition}

Note that for every $ 1 \le i \le |Q| $ we have $ A_{\min}[i] \le i \le A_{\max}[i] $ because there exists an $ \epsilon $-walk from $ Q[i] $ to $ Q[i] $. In Figure \ref{fig:example} we have $ A_{\max}[3] = A_{\max}[4] = 5 $ and $ A_{\max}[i] = i $ for $ i \not \in \{3, 4 \} $.

We will use $ A_{\max} $ and $ A_{\min} $ to prove Theorem \ref{theor:fmindex}, but we will not explicitly store $ A_{\max} $ and $ A_{\min} $ (which would require $ |Q| \log |Q| $ bits per array), because otherwise we could not achieve the space bound of $ \mathfrak{e} \log \sigma (1 + o(1)) + O(e) $ bits that holds for Wheeler GNFAs without $ \epsilon $-transitions.

Consider again Figure \ref{fig:example}. We have observed that, for $ \alpha = cba $, the largest integer $ 0 \le j \le |Q| $ such that the two conditions of Lemma \ref{lem:lessold} are true is $ 4 $. Intuitively, since $ A_{\max}[4] > 4 $, then we must have $ |G^\prec(\alpha)| \le 3 $. We also know that $ A_{\max}[3] > 3 $, so we must have $ |G^\prec(\alpha)| \le 2 $. Since $ A_{\max}[2] = 2 $, then we can conclude $ |G^\prec(\alpha)| = 2 $. We can now formally prove an extension of Lemma \ref{lem:lessold} to arbitrary Wheeler GNFAs.

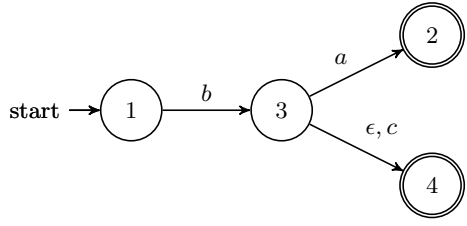
\begin{figure}[h]
     \centering
        \scalebox{1}{
        \begin{tikzpicture}[->,>=stealth', semithick, auto, scale=1]
%\tikzset{every state/.style={minimum size=0pt}}
\node[state, initial, initial] (1)    at (0,0)	{$ 1 $};
\node[state] (3)    at (2, 0)	{$ 3 $};
\node[state, accepting] (2)    at (4, 1)	{$ 2 $};
\node[state, accepting] (4)    at (4, -1)	{$ 4 $};

\draw (1) edge [] node [] {$ b $} (3);
\draw (3) edge [] node [] {$ a $} (2);
\draw (3) edge [] node [] {$ \epsilon, c $} (4);

\end{tikzpicture}
}
\caption{A Wheeler $ 1 $-GNFA showing that in Lemma \ref{lem:lessequalnew} we need to distinguish between the case $ h^* = |G^\prec(\alpha)| $ and the case $ h^* > |G^\prec(\alpha)| $. The states are numbered following the Wheeler order.}
\label{fig:twocasesneeded}
\end{figure}

\begin{lemma}\label{lem:lessnew}
    Let $ \mathcal{A} = (Q, E, s, F) $ be a Wheeler $ r $-GNFA, let $ \le $ be a Wheeler order on $ \mathcal{A} $, and let $ \alpha \in \Sigma^*$, with $ \alpha \not = \epsilon $. For $ 0 < k < \min\{r + 1,|\alpha|\} $, let $ f_k = \mathtt{out}(Q[1, |G^\prec(p(\alpha, |\alpha| - k))|], s(\alpha, k)) $. Let $ j^* $ be the largest integer $ 0 \le j \le |Q| $ such that:
    \begin{itemize}
        \item $ \mathtt{in}(Q[1, j], s(\alpha, k)) \leq f_k $, for every $ 0 < k < \min\{r + 1,|\alpha|\} $;
        \item $ \rho \prec \alpha $ for every $ \rho \in \Sigma^k \cap \lambda (Q[h]) $, for every $ 1 \le h \le j $ and for every $ 0 < k < r + 1 $.
    \end{itemize}
    Then, $ |G^\prec(\alpha)| $ is the largest integer $ 0 \le t \le j^* $ such that, if $ t \ge 1 $, then $ A_{\max}[t] = t $.
\end{lemma}

Similarly, we will now prove an extension of Lemma \ref{lem:lessequalold} to arbitrary Wheeler GNFAs. We will now use the array $ A_{\min} $ (and not $ A_{\max} $). Intuitively, Lemma \ref{lem:lessequalold} could return a value of $ |G^\prec_\dashv(\alpha)| $ \emph{smaller} (and not larger) than the correct value, so we will use $ A_{\min} $ to determine $ |G^\prec_\dashv(\alpha)| $. However, we must be careful. In Lemma \ref{lem:lessequalnew}, we will introduce the integer $ h^* $ and we will distinguish between the case $ h^* = |G^\prec(\alpha)| $ and the case $ h^* > |G^\prec(\alpha)| $. If $ h^* = |G^\prec(\alpha)| $, we can prove that $ G_\dashv(\alpha) = \emptyset $ and so we must have $ |G^\prec_\dashv(\alpha)| = h^* = |G^\prec(\alpha)| $. If $ h^* > |G^\prec(\alpha)| $, then we can proceed similarly to Lemma \ref{lem:lessnew}. Notice that we cannot collapse the two cases into a single case by always returning the smallest integer $ h^* \le t \le |Q| $ such that, if $ t < |Q| $, then $ A_{\min}[t + 1] = t + 1 $. In Figure \ref{fig:twocasesneeded}, for $ \alpha = bb $, we have $ h^* = |G^\prec(\alpha)| = 3 = |G^\prec_\dashv(\alpha)| $ and the smallest integer $ h^* \le t \le |Q| $ such that, if $ t < |Q| $, then $ A_{\min}[t + 1] = t + 1 $ is equal to $ 4 $, not $ 3 $.

\begin{lemma}\label{lem:lessequalnew}
Let $ \mathcal{A} = (Q, E, s, F) $ be a Wheeler $ r $-GNFA, let $ \le $ be a Wheeler order on $ \mathcal{A} $, and let $\alpha \in \Sigma^*$, with $ \alpha \not = \epsilon $. For $ 0 < k < \min\{r + 1,|\alpha|\} $, let $ f_k = \mathtt{out}(Q[1, |G^\prec(p(\alpha, |\alpha| - k))|], s(\alpha, k)) $ and let $ g_k = \mathtt{out}(Q[1, |G^\prec_\dashv(p(\alpha, |\alpha| - k))|], s(\alpha, k)) $. Then, $ g_k \ge f_k $ for every $ 0 < k < \min\{r + 1,|\alpha|\} $. Let $ h^* $ be the maximum among:
\begin{itemize}
    \item $ |G^\prec(\alpha)| $;
    \item the largest integer $ 0 \le i \le |Q| $ such that, if $ i \ge 1 $, then $ Q[i] \in G^* (\alpha) $;
    \item the smallest integer $ 0 \le j \le |Q| $ such that, for every $ 0 < k < \min\{r + 1,|\alpha|\} $ for which $ g_k > f_k $, we have $\mathtt{in}(Q[1, j], s(\alpha, k)) \ge g_k $.
\end{itemize}
Then, $ h^* \ge |G^\prec(\alpha)| $. Moreover:
\begin{itemize}
    \item If $ h^* = |G^\prec(\alpha)| $, then $ |G^\prec_\dashv(\alpha)| = h^* = |G^\prec(\alpha)| $.
    \item If $ h^* > |G^\prec(\alpha)| $, then $ |G^\prec_\dashv(\alpha)| \ge h^* > |G^\prec(\alpha)| $, and $ |G^\prec_\dashv(\alpha)| $ is the smallest integer $ h^* \le t \le |Q| $ such that, if $ t < |Q| $, then $ A_{\min}[t + 1] = t + 1 $.
\end{itemize}
\end{lemma}

Given a bit array $ B[1, n] $, we can define the following queries: (i) $ \access (B, i) $: given $ 1 \le i \le n $, return $ B[i] $; (ii) $ \rank_1 (B, i) $: given $ 1 \le i \le n $, return $ |\{1 \le h \le i \;|\; B[h] = 1 \}| $; (iii) $ \select_1 (B, i) $: given $ 1 \le i \le  \rank_1 (B, n) $, return the unique $ 1 \le j \le n $ such that $ B[j] = 1 $ and $ |\{1 \le h \le j \;|\; B[h] = 1 \}| = i $. A bit array $ B $ of length $ n $ can be stored using a data structure (called \emph{bitvector}) of $ n + o(n) $ bits that supports $ \access $, $ \rank $ and $ \select $ queries in $ O(1) $ time and can be built from $ B $ in $ O(n) $ time \cite{navarro2016}.

We are now ready to prove Theorem \ref{theor:fmindex} for arbitrary Wheeler GNFAs. We store the same data structure of $ \mathfrak{e} \log \sigma (1 + o(1)) + O(e) $ bits used in \cite{cotumaccio2024stacs} for Wheeler GNFAs without $ \epsilon $-transitions. Moreover, let $ B_{\max}[1, |Q|] $ ($ B_{\min}[1, |Q|] $, respectively) the bit array such, for every $ 1 \le i \le |Q| $, we have $ B_{\max}[i] = 1 $ ($ B_{\min}[i] = 1 $, respectively) if and only if $ A_{\max}[i] = i $ ($ A_{\min}[i] = i $, respectively). We can store $ B_{\max} $ and $ B_{\min} $ as bitvectors using $ 2|Q| + o(|Q|) $ bits, which is absorbed by the $ O(e) $ term (we have $ e \ge |Q| - 1 $ because every state is reachable from the initial state). To compute $ G_\dashv(\alpha) $, we proceed as explained in Section \ref{sec:howtosolve}, but we use Lemma \ref{lem:lessnew} and Lemma \ref{lem:lessequalnew} instead of Lemma \ref{lem:lessold} and Lemma \ref{lem:lessequalold}.
\begin{itemize}
    \item To compute $ |G^\prec(\alpha)| $, we first compute the value $ j^* $ of Lemma \ref{lem:lessnew} by proceeding as in Lemma \ref{lem:lessold}. By Lemma \ref{lem:lessnew}, if $ \rank_1 (B_{\max}, j^*) = 0 $, we conclude $ |G^\prec(\alpha)| = 0 $; otherwise, we conclude $ |G^\prec(\alpha)| = \select_1 (B_{\max}, \rank_1 (B_{\max}, j^*)) $.
    \item To compute $ |G^\prec_\dashv(\alpha)| $, we first compute the value $ h^* $ of Lemma \ref{lem:lessequalnew} by proceeding as in Lemma \ref{lem:lessequalold}. By Lemma \ref{lem:lessequalnew}, if $ h^* = |G^\prec(\alpha)| $, then we conclude $ |G^\prec_\dashv(\alpha)| = h^* $. Now assume that $ h^* > |G^\prec(\alpha)| $. If $ \rank_1 (B_{\min}, |Q|) = \rank_1 (B_{\min}, h^*) $, then $ |G^\prec_\dashv(\alpha)| = |Q| $; otherwise, we conclude $ |G^\prec_\dashv(\alpha)| = \select_1 (B_{\min}, \rank_1 (B_{\min}, h^*) + 1) - 1 $.
\end{itemize}

\section{Construction}\label{sec:construction}

In Section \ref{sec:patternepsilon}, we showed that by storing a bitvector representation of $ B_{\max} $ and $ B_{\min} $ we can prove Theorem \ref{theor:fmindex} for an arbitrary Wheeler GNFA $ \mathcal{A} = (Q, E, s, F) $. Let us show that we can build $ B_{\max} $ and $ B_{\min} $ in $ O(e) $ time, where $ e = |E| $ (recall that $ e \ge |Q| - 1 $). We focus on $ B_{\max} $ because the proof for $ B_{\min} $ is analogous. For every $ 1 \le i \le |Q| $, we have $ B_{\max}[i] = 1 $ if and only if $ A_{\max}[i] = i $, so we only have to show how to build $ A_{\max} $ in $ O(e) $ time.

Let $ \mathcal{A} = (Q, E, s, F) $ be a Wheeler GNFA, with Wheeler order $ \le $. We have the following properties.
\begin{itemize}
    \item \emph{(Property 1)} For every $ t \ge 2 $, there exist no pairwise distinct $ u_1, u_2, \dots, u_t \in Q $ such that $ (u_i, u_{i + 1}, \epsilon) \in E $ for every $ 1 \le i \le t - 1 $ and $ (u_t, u_1, \epsilon) $. Indeed, suppose for the sake of a contradiction that such $ u_1, u_2, \dots, u_t \in Q $ exist. In particular, $ u_1 \not = u_2 $. Suppose that $ u_1 < u_2 $ (the case $ u_2 < u_1 $ leads to a contradiction analogously). Since $ (u_t, u_1, \epsilon), (u_1, u_2, \epsilon) \in E $, from Axiom 4 we obtain $ u_t \le u_1 $, and we have $ u_t < u_1 $ because the $ u_i $'s are pairwise distinct. Since $ (u_{t - 1}, u_t, \epsilon), (u_t, u_1, \epsilon) \in E $ , we analogously obtain $ u_{t - 1} < u_t $. By iterating the same argument, we obtain $ u_{t - 2} < u_{t - 1} $, $ \dots $, $ u_2 < u_3 $. We conclude $ u_2 < u_3 < \dots < u_{t - 2} < u_{t - 1} < u_t < u_1 < u_2 $ and so $ u_2 < u_2 $, a contradiction.
    \item \emph{(Property 2)} Let us prove that for every $ 1 \le i \le |Q| $, we have:
    \begin{equation*}
        A_{\max}[i] = \max \{i, \max \{A_{\max}[j] \;|\; 1 \le j \le |Q|, j \not = i , (Q[j], Q[i], \epsilon) \in E \} \}.
    \end{equation*}
    $(\ge) $ First, we have $ A_{\max}[i] \ge i $ because there exists an $ \epsilon $-walk from $ Q[i] $ to $ Q[i] $. Next, for every $ 1 \le j \le |Q| $ such that $ j \not = i $ and $ (Q[j], Q[i], \epsilon) \in E $, we have $ A_{\max}[i] \ge A_{\max}[j] $ because there exists an $ \epsilon $-walk from $ Q[A_{\max}[j]] $ to $ Q[j] $, so there exists an $ \epsilon $-walk from $ Q[A_{\max}[j]] $ to $ Q[i] $.

    $ (\le) $ If $ A_{\max}[i] = i $, we are done. Now assume that $ A_{\max}[i] \not = i $. We know that there exists an $ \epsilon $-walk from $ Q[A_{\max}[i]] $ to $ Q[i] $, so there exists $ 1 \le j \le Q $ such that $ j \not = i $, $ (Q[j], Q[i], \epsilon) \in E $ and there exists an $ \epsilon $-walk from $ Q[A_{\max}[i]] $ to $ Q[j] $, which implies $ A_{\max}[i] \le A_{\max}[j] $.
\end{itemize}

We are now ready to prove the main result of this section.

\begin{theorem}\label{theor:construction}
    Let $ \mathcal{A} = (Q, E, s, F) $ be a Wheeler GNFA. Then, in $ O(e) $ time we can build $ A_{\max} $ and $ A_{\min} $.
\end{theorem}

\begin{proof}
    We present only the proof for $ A_{\max} $ because the proof for $ A_{\min} $ is analogous. Consider an array $ A[1, |Q|] $. At the beginning of the algorithm, we have $ A[i] = i $ for every $ 1 \le i \le |Q| $, and at the end of the algorithm we will have $ A[i] = A_{\max}[i] $ for every $ 1 \le i \le |Q| $. We perform a depth-first search on the transpose of $ \mathcal{A} $ (that is, we follow the edges in a backward fashion). In other words, when we process a node $ u $, we consider all nodes $ v $ such that for some $ \rho \in \Sigma^* $ we have $ (v, u, \rho) \in E $. At any time, every node is white, gray or black. A white node can only become gray and a gray node can only become black. Every node is white at the beginning of the algorithm, and every node is black at the end of the algorithm.
    
    The algorithm considers all nodes $ Q[i] $. If $ Q[i] $ is white, we paint $ Q[i] $ gray and we consider all $ j $'s such that $ j \not = i $ and $ (Q[j], Q[i], \epsilon) \in E $. For every such node $ Q[j] $, if $ Q[j] $ is black and $ A[j] > A[i] $, we set $ A[i] \gets A[j] $, and if $ Q[j] $ is white, we recursively process $ Q[j] $ and when the recursion is complete, if $ A[j] > A[i] $, we set $ A[i] \gets A[j] $. After processing every $ j $, we paint $ Q[i] $ black and we process the next $ Q[i] $.

    The algorithm runs in $ O(e) $ time because every edge is processed once. Note that after a node $ Q[i] $ becomes black, the value $ A[i] $ is no longer updated. Moreover, when we consider all $ j $'s such that $ j \not = i $ and $ (Q[j], Q[i], \epsilon) \in E $, no such $ j $ can be gray by Property 1. As a consequence, by Property 2 we conclude that at the end of the algorithm we have $ A[i] = A_{\max}[i] $ for every $ 1 \le i \le |Q| $, which can be seen by considering all $ i $'s in the order in which they become black. \hfill $ \qed $
\end{proof}

\section{Conclusions and Future Work}\label{sec:conclusions}

In this paper, we have shown how to achieve linear-time pattern matching on arbitrary Wheeler GNFAs within compressed space (Theorem \ref{theor:fmindex}). To this end, we have introduced two arrays $ A_{\max} $ and $ A_{\min} $ that can be built in linear time (Theorem \ref{theor:construction}). Our result subsumes all previous approaches based on the Burrows-Wheeler transform and the FM-index, including labeled trees \cite{ferraginajacm2009}, de Bruijn graphs \cite{bowe2012} and conventional Wheeler graphs \cite{gagie2017,alanko2020}, which have led to a fruitful line of research \cite{alanko2022linear,conte2023computing,cotumaccio2023space,cotumaccio2023prefix,cotumaccio2024convex,alanko2024computing}.

There are several potential directions for future research. The ideas behind Wheeler NFAs have been extended to arbitrary NFAs \cite{cotumaccio2021,cotumaccio2022}, so the next natural step is to extend the same paradigm to arbitrary GNFAs (however, on arbitrary NFAs and GNFAs we cannot achieve linear-time pattern matching because of Equi et al.'s lower bound mentioned in the introduction). It is still an open problem to characterize the class of all regular languages recognized by some Wheeler GNFA, even though it is known that this class is strictly larger than the class of all regular languages recognized by some Wheeler NFA \cite{cotumaccio2024stacs}. Moreover, in this paper we focused on $ \epsilon $-transitions, so it should be possible to extend the results in \cite{cotumacciojacm} on the complexity of regular expression by leveraging Thompson's construction for converting a regular expression into an equivalent automaton \cite{hopcroft2006}, which heavily relies on $ \epsilon $-transitions.

\begin{credits}
\subsubsection{\ackname} Funded by the Helsinki Institute for Information Technology (HIIT).

%\subsubsection{\discintname}
%It is now necessary to declare any competing interests or to specifically
%state that the authors have no competing interests. Please place the
%statement with a bold run-in heading in small font size beneath the
%(optional) acknowledgments\footnote{If EquinOCS, our proceedings submission
%system, is used, then the disclaimer can be provided directly in the system.},
%for example: The authors have no competing interests to declare that are
%relevant to the content of this article. Or: Author A has received research
%grants from Company W. Author B has received a speaker honorarium from
%Company X and owns stock in Company Y. Author C is a member of committee Z.
\end{credits}


\bibliographystyle{splncs04}
\bibliography{mybibliography}

\newpage

\appendix

\section{Omitted Proofs}

\noindent{\textbf{Statement of Lemma \ref{lem:intervals}}.}
Let $ \mathcal{A} = (Q, E, s, F) $ be a Wheeler GNFA, let $ \le $ be a Wheeler order on $ \mathcal{A} $, and let $ \alpha \in \Sigma^* $. Then:
\begin{enumerate}
    \item $ G^\prec(\alpha) \cap G_\dashv(\alpha) = \emptyset $.
    \item $ G_\dashv(\alpha) $ is $ \le $-convex.
    \item If $ u, v \in Q $ are such that $ u < v $ and $ v \in G^\prec(\alpha) $, then $ u \in G^\prec(\alpha) $. In other words, $ G^\prec(\alpha) = Q[1, |G^\prec(\alpha)|] $.
    \item If $ u, v \in Q $ are such that $ u < v $ and $ v \in G^\prec_\dashv(\alpha) $, then $ u \in G^\prec_\dashv(\alpha) $. In other words, $ G^\prec_\dashv(\alpha) = Q[1, |G^\prec_\dashv(\alpha)|] $.
    \item $ G_\dashv(\alpha) = Q[|G^\prec(\alpha)| + 1, |G^\prec_\dashv(\alpha)|] $.
\end{enumerate}

\begin{proof}
\begin{enumerate}
    \item If $ u \in G_\dashv(\alpha) $, then there exists $ \beta \in I_u $ such that $ \alpha \dashv \beta $. In particular, $ \alpha \preceq \beta $, so $ u \not \in G^\prec(\alpha) $.
    \item Assume that $ u, v, z \in Q $ are such that $ u < v < z $ and $ u, z \in G_\dashv(\alpha) $. We must prove that $ v \in G_\dashv(\alpha) $. Since $ u, z \in G_\dashv(\alpha) $, then there exist $ \beta \in I_u $ and $ \delta \in I_z $ such that $ \alpha \dashv \beta $ and $ \alpha \dashv \delta $. If $ (\beta \in I_v) \lor (\delta \in I_v) $ the conclusion follows, so in the following we assume $ \beta \not \in I_v $ and $ \delta \not \in I_v $. Fix any $ \gamma \in I_v $; we only have to prove that $ \alpha \dashv \gamma $. From $ u < v < z $, we obtain $ u \preceq_\mathcal{A} v \preceq_\mathcal{A} z $ by Axiom 1. Since $ \beta \in I_u $, $ \gamma \in I_v $ and $ \{\beta, \gamma \} \not \subseteq I_u \cap I_v $, we conclude $ \beta \prec \gamma $. Since $ \gamma \in I_v $, $ \delta \in I_z $ and $ \{\gamma, \delta \} \not \subseteq I_v \cap I_z $, we conclude $ \gamma \prec \delta $. Hence, we obtain $ \beta \prec \gamma \prec \delta $, and from $ \alpha \dashv \beta $ and $ \alpha \dashv \delta $ we conclude $ \alpha \dashv \gamma $.
    \item Let $ \beta \in I_u $. We must prove that $ \beta \prec \alpha $. If $ \beta \in I_v $, then from $ v \in G^\prec(\alpha) $ we conclude $ \beta \prec \alpha $, so in the following we assume $ \beta \not \in I_v $. Fix any $ \gamma \in I_v $. Since $ u < v $, then we have $ u \preceq_\mathcal{A} v $ by Axiom 1. From $ \beta \in I_u $, $ \gamma \in I_v $ and $ \{\beta, \gamma \} \not \subseteq I_u \cap I_v $ we conclude $ \beta \prec \gamma $. From $ v \in G^\prec(\alpha) $ we obtain $ \gamma \prec \alpha $, so we conclude $ \beta \prec \alpha $.
    \item Since $ v \in G^\prec_\dashv(\alpha) $ , we have either $ v \in G^\prec(\alpha) $ or $ v \in G_\dashv(\alpha) $. If $ v \in G^\prec(\alpha) $, then $ u \in G^\prec(\alpha) $ by the previous point and so $ u \in G^\prec_\dashv(\alpha) $. Now assume that $ v \in G_\dashv(\alpha) $. If $ u \in G_\dashv(\alpha) $, then $ u \in G^\prec_\dashv(\alpha) $ and we are done, so we can assume $ u \not \in G_\dashv(\alpha) $. Let us prove that it must be $ u \in G^\prec(\alpha) $, which again implies $ u \in G^\prec_\dashv(\alpha) $. Fix $ \beta \in I_u $; we must prove that $ \beta \prec \alpha $. Since $ v \in G_\dashv(\alpha) $, then there exists $ \gamma \in I_v $ such that $ \alpha \dashv \gamma $. From $ u \not \in G_\dashv(\alpha) $ we obtain $ \gamma \not \in I_u $. From $ u < v $ we obtain $ u \preceq_\mathcal{A} v $ by Axiom 1, so from $ \beta \in I_u $, $ \gamma \in I_v $ and $ \{\beta, \gamma \} \not \subseteq I_u \cap I_v $ we conclude $ \beta \prec \gamma $. Since $ u \not \in G_\dashv(\alpha) $ we have $ \lnot (\alpha \dashv \beta) $, so from $ \beta \prec \gamma $ and $ \alpha \dashv \gamma $ we conclude $\beta \prec \alpha $.
    \item By the definition of $ G^\prec_\dashv(\alpha) $ and the first point we obtain that $ G^\prec_\dashv(\alpha) $ is the disjoint union of $ G^\prec(\alpha) $ and $ G_\dashv(\alpha) $, so the conclusion follows from the third point and the fourth point. \hfill $ \qed $
\end{enumerate}
\end{proof}

\noindent{\textbf{Statement of Lemma \ref{lem:lessnew}}.}
    Let $ \mathcal{A} = (Q, E, s, F) $ be a Wheeler $ r $-GNFA, let $ \le $ be a Wheeler order on $ \mathcal{A} $, and let $ \alpha \in \Sigma^*$, with $ \alpha \not = \epsilon $. For $ 0 < k < \min\{r + 1,|\alpha|\} $, let $ f_k = \mathtt{out}(Q[1, |G^\prec(p(\alpha, |\alpha| - k))|], s(\alpha, k)) $. Let $ j^* $ be the largest integer $ 0 \le j \le |Q| $ such that:
    \begin{itemize}
        \item $ \mathtt{in}(Q[1, j], s(\alpha, k)) \leq f_k $, for every $ 0 < k < \min\{r + 1,|\alpha|\} $;
        \item $ \rho \prec \alpha $ for every $ \rho \in \Sigma^k \cap \lambda (Q[h]) $, for every $ 1 \le h \le j $ and for every $ 0 < k < r + 1 $.
    \end{itemize}
    Then, $ |G^\prec(\alpha)| $ is the largest integer $ 0 \le t \le j^* $ such that, if $ t \ge 1 $, then $ A_{\max}[t] = t $.

\begin{proof}
    First, let us prove that $ |G^\prec(\alpha)| \le j^* $. The conclusion is immediate if $ j^* = |Q| $, so we can assume $ j^* \le |Q| - 1 $. We only have to prove that $ Q[j^* + 1] \not \in G^\prec(\alpha) $. By the maximality of $ j^* $, we know that at least one of the following is true:
    \begin{enumerate}
        \item There exists $ 0 < k < \min\{r + 1,|\alpha|\} $ such that $ \mathtt{in}(Q[1, j^* + 1], s(\alpha, k)) > f_k $.
        \item There exist $ 0 < k < r + 1 $, $ 1 \le h \le j^* + 1 $ and $ \rho \in \Sigma^k \cap \lambda (Q[h]) $ such that $ \alpha \preceq \rho $.
    \end{enumerate}
    We consider the two cases separately.
    \begin{enumerate}
        \item Let $ 0 < k < \min\{r + 1,|\alpha|\} $ be such that $ \mathtt{in}(Q[1, j^* + 1], s(\alpha, k)) > f_k $. The conclusion will follow if we show that there exists $ u' \in Q $ such that $ (u', Q[j^* + 1], s(\alpha, k)) \in E $ and $ u' \not \in G^\prec(p(\alpha, |\alpha| - k)) $. Indeed, suppose that we have proved that such an $ u' $ exists. Since $ u' \not \in G^\prec(p(\alpha, |\alpha| - k)) $, then there exists $ \beta \in I_{u'} $ such that $ p(\alpha, |\alpha| - k) \preceq \beta $, so $ \beta s(\alpha, k) \in I_{Q[j^* + 1]} $ and $ \alpha = p(\alpha, |\alpha| - k) s(\alpha, k) \preceq \beta s(\alpha, k) $, which proves that $ Q[j^* + 1] \not \in G^\prec(\alpha) $.

        From the definition of $ j^* $ we know that $ \mathtt{in}(Q[1, j^*], s(\alpha, k)) \le f_k $, so from $ \mathtt{in}(Q[1, j^* + 1], s(\alpha, k)) > f_k $ we conclude that there exists $ u^* \in Q $ such that $ (u^*, Q[j^* + 1], s(\alpha, k)) \in E $. Let $ u' $ be the largest (w.r.t $ \le $) state in $ Q $  such that $ (u', Q[j^* + 1], s(\alpha, k)) \in E $. We only have to prove that $ u' \not \in G^\prec(p(\alpha, |\alpha| - k)) $.
        
        Suppose for the sake of a contradiction that $ u' \in G^\prec(p(\alpha, |\alpha| - k)) $. We will obtain a contradiction by showing that $ \mathtt{in}(Q[1, j^* + 1], s(\alpha, k)) \le f_k $. To this end, we only have to show that for every $ (u, v, s(\alpha, k)) \in E $, if $ v \le Q[j^* + 1] $, then $ u \in G^\prec(p(\alpha, |\alpha| - k)) $. Since $ u' \in G^\prec(p(\alpha, |\alpha| - k)) $, the conclusion will follow if we prove that for every $ (u, v, s(\alpha, k)) \in E $, if $ v \le Q[j^* + 1] $, then $ u \le u' $. We distinguish two cases.
        \begin{itemize}
            \item Assume that $ v < Q[j^* + 1] $. If we consider $ (u, v, s(\alpha, k)), (u', Q[j^* + 1], s(\alpha, k)) \in E $, then by Axiom 4 we conclude $ u \le u' $.
            \item Assume that $ v = Q[j^* + 1] $. Then, by the maximality of $ u' $ we conclude $ u \le u' $.
        \end{itemize}
    \item Let $ 0 < k < r + 1 $, $ 1 \le h \le j^* + 1 $ and $ \rho \in \Sigma^k \cap \lambda (Q[h]) $ be such that $ \alpha \preceq \rho $. Then, we must have $ h = j^* + 1 $, because if we had $ h \le j^* $, then from the definition of $ j^* $ we would conclude $ \rho \prec \alpha $. In other words, we have $ \rho \in \Sigma^k \cap \lambda (Q[j^* + 1]) $. Since $ \rho \in \lambda (Q[j^* + 1]) $, then there exists $ u' \in Q $ such that $ (u', Q[j^* + 1], \rho) \in E $. Pick any $ \beta \in I_{u'} $. Then, $ \beta \rho \in I_{Q[j^* + 1]} $. From $ \alpha \preceq \rho $ we conclude $ \alpha \preceq \beta \rho $, so $ Q[j^* + 1] \not \in G^\prec(\alpha) $.
    \end{enumerate}

Now, let $ t^* $ be the largest integer $ 0 \le t \le j^* $ such that, if $ t \ge 1 $, then $ A_{\max}[t] = t $. We need to prove that $ |G^\prec(\alpha)| = t^* $

$ (\le) $ Let us prove that for every $ 0 \le d \le j^* - t^* $ we have $ |G^\prec(\alpha)| \le j^* - d $. The conclusion will follow by choosing $ d = j^* - t^* $. We proceed by induction on $ d $. If $ d = 0 $, we know that $ |G^\prec(\alpha)| \le j^* $. Now, assume that $ 1 \le d \le j^* - t^* $ (which implies $ 1 \le j^* - d + 1 \le |Q| $ because $ 1 \le t^* + 1 \le j^* - d + 1 \le j^* \le |Q| $). To prove that $ |G^\prec(\alpha)| \le j^* - d $, it will suffice to prove that $ Q[j^* - d + 1] \not \in G^\prec (\alpha) $. Notice that $ t^* + 1 \le j^* - d + 1 \le j^* $, so by the maximality of $ t^* $ we have $ A_{\max}[j^* - d + 1] > j^* - d + 1 $. This means that for some $ j^* - d + 2 \le j \le |Q| $ there exists an $ \epsilon $-walk from $ Q[j] $ to $ Q[j^* - d + 1] $. By the inductive hypothesis, we know that $ |G^\prec(\alpha)| \le j^* - d + 1 $, so $ Q[j] \not \in G^\prec(\alpha) $ and there exists $ \beta \in I_{Q[j]} $ such that $ \alpha \preceq \beta $. Since there exists an $ \epsilon $-walk from $ Q[j] $ to $ Q[j^* - d + 1] $, we conclude that $ \beta \in I_{Q[j^* - d + 1]} $, so $ Q[j^* - d + 1] \not \in G^\prec(\alpha) $ because $ \alpha \preceq \beta $.

$ (\ge) $ If $ t^* = 0 $ the conclusion is immediate, so we can assume $ t^* \ge 1 $. We only need to show that $ Q[t^*] \in G^\prec(\alpha) $. Fix any $ \beta \in I_{Q[t^*]} $. We must prove that $ \beta \prec \alpha $. If $ \beta = \epsilon $ the conclusion is immediate (because $ \alpha \not = \epsilon $), so we can assume $ \beta \not = \epsilon $. This implies that there exist $ 1 \le g \le |Q| $, $ u^* \in Q $ and $ 0 < k \le \min \{r, |\beta| \} $ such that (i) there exists an $ \epsilon $-walk from $ Q[g] $ to $ Q[t^*] $, (ii) $ (u^*, Q[g], s(\beta, k)) \in E $ and (iii) $ p(\beta, |\beta| - k) \in I_{u^*} $. By the definition of $ t^* $ we know that $ A_{\max}[t^*] = t^* $, so $ g \le t^* $. From $ t^* \le j^* $ we obtain $ g \le j^* $, and from $ (u^*, Q[g], s(\beta, k)) \in E $ we obtain $ s(\beta, k) \in \Sigma^k \cap \lambda (Q[g]) $, so from the definition of $ j^* $ we conclude $ s(\beta, k) \prec \alpha $. We now distinguish two cases.
\begin{enumerate}
    \item Assume that $ \lnot(s(\beta, k) \dashv \alpha) $. Then, from $ s(\beta, k) \prec \alpha $ we obtain $ \beta \prec \alpha $.
    \item Assume that $ s(\beta, k) \dashv \alpha $. Then, $ s(\beta, k) = s(\alpha, k) $, and from $ s(\beta, k) \prec \alpha $ we obtain $ k < |\alpha| $, so that $ 0 < k < \min \{r + 1, |\alpha| \} $. The conclusion will follow if we prove that $ u^* \in G^\prec(p(\alpha, |\alpha| - k)) $, because from $ p(\beta, |\beta| - k) \in I_{u^*} $ we will conclude $ p(\beta, |\beta| - k) \prec p(\alpha, |\alpha| - k) $ and so $ \beta = p(\beta, |\beta| - k) s(\beta, k) = p(\beta, |\beta| - k) s(\alpha, k) \prec p(\alpha, |\alpha| - k) s(\alpha, k) = \alpha $. Note that, since $ g \le j^* $, then from the definition of $ j^* $ we obtain $ \mathtt{in}(Q[1, g], s(\alpha, k)) \le \mathtt{in}(Q[1, j^*], s(\alpha, k)) \leq f_k $.

    Suppose for the sake of a contradiction that $ u^* \not \in G^\prec(p(\alpha, |\alpha| - k)) $. We will obtain a contradiction by showing that $ \mathtt{out}(Q[1, |G^\prec(p(\alpha, |\alpha| - k))|], s(\alpha, k)) < f_k $. To this end, we only need to show that for every $ (u, v, s(\alpha, k)) \in E $, if $ u \in G^\prec(p(\alpha, |\alpha| - k)) $, then $ v \le Q[g] $, because $ \mathtt{in}(Q[1, g], s(\alpha, k)) \le f_k $, $ (u^*, Q[g], s(\alpha, k)) = (u^*, Q[g], s(\beta, k)) \in E $ and $ u^* \not \in G^\prec(p(\alpha, |\alpha| - k)) $.

     Fix $ (u, v, s(\alpha, k)) \in E $ such that $ u \in G^\prec(p(\alpha, |\alpha| - k)) $. We must prove that $ v \le Q[g] $. Suppose for the sake of a contradiction that $ Q[g] < v $. If we consider $ (u^*, Q[g], s(\alpha, k)), (u, v, s(\alpha, k)) \in E $, then by Axiom 4 we conclude $ u^* \le u $, so from $ u \in G^\prec(p(\alpha, |\alpha| - k)) $ we obtain $ u^* \in G^\prec(p(\alpha, |\alpha| - k)) $, a contradiction. \hfill $ \qed $
\end{enumerate}
\end{proof}

\noindent{\textbf{Statement of Lemma \ref{lem:lessequalnew}}.}
Let $ \mathcal{A} = (Q, E, s, F) $ be a Wheeler $ r $-GNFA, let $ \le $ be a Wheeler order on $ \mathcal{A} $, and let $\alpha \in \Sigma^*$, with $ \alpha \not = \epsilon $. For $ 0 < k < \min\{r + 1,|\alpha|\} $, let $ f_k = \mathtt{out}(Q[1, |G^\prec(p(\alpha, |\alpha| - k))|], s(\alpha, k)) $ and let $ g_k = \mathtt{out}(Q[1, |G^\prec_\dashv(p(\alpha, |\alpha| - k))|], s(\alpha, k)) $. Then, $ g_k \ge f_k $ for every $ 0 < k < \min\{r + 1,|\alpha|\} $. Let $ h^* $ be the maximum among:
\begin{itemize}
    \item $ |G^\prec(\alpha)| $;
    \item the largest integer $ 0 \le i \le |Q| $ such that, if $ i \ge 1 $, then $ Q[i] \in G^* (\alpha) $;
    \item the smallest integer $ 0 \le j \le |Q| $ such that, for every $ 0 < k < \min\{r + 1,|\alpha|\} $ for which $ g_k > f_k $, we have $\mathtt{in}(Q[1, j], s(\alpha, k)) \ge g_k $.
\end{itemize}
Then, $ h^* \ge |G^\prec(\alpha)| $. Moreover:
\begin{itemize}
    \item If $ h^* = |G^\prec(\alpha)| $, then $ |G^\prec_\dashv(\alpha)| = h^* = |G^\prec(\alpha)| $.
    \item If $ h^* > |G^\prec(\alpha)| $, then $ |G^\prec_\dashv(\alpha)| \ge h^* > |G^\prec(\alpha)| $, and $ |G^\prec_\dashv(\alpha)| $ is the smallest integer $ h^* \le t \le |Q| $ such that, if $ t < |Q| $, then $ A_{\min}[t + 1] = t + 1 $.
\end{itemize}

\begin{proof}
    For every $ 0 < k < \min\{r + 1,|\alpha|\} $ we have $ g_k \ge f_k $ because $ G^\prec(p(\alpha, |\alpha| - k) \subseteq G^\prec_\dashv(p(\alpha, |\alpha| - k) $.
    
    Let $ i^* $ be the largest integer $ 0 \le i \le |Q| $ such that, if $ i \ge 1 $, then $ Q[i] \in G^* (\alpha) $, and let $ j^* $ be the smallest integer $ 0 \le j \le |Q| $ such that $\mathtt{in}(Q[1, j], s(\alpha, k)) \ge g_k $ for every $ 0 < k < \min\{r + 1,|\alpha|\} $ such that $ g_k > f_k $. By definition, we have $ h^* = \max \{|G^\prec(\alpha)|, i^*, j^* \} $, and in particular $ h^* \ge |G^\prec(\alpha)| $.

    Let $ t^* $ be the smallest integer $ h^* \le t \le |Q| $ such that, if $ t < |Q| $, then $ A_{\min}[t + 1] = t + 1 $.

    We will prove the following statements:
    \begin{enumerate}
        \item If $ i^* \ge 1 $, then $ Q[i^*] \in G_\dashv (\alpha) $.
        \item If $ j^* \ge 1 $, then $ Q[j^*] \in G_\dashv (\alpha) $.
        \item If $ h^* = |G^\prec(\alpha)| $, then $ i^* = j^* = 0 $.
        \item If there exists $ 0 < k < \min\{r + 1,|\alpha|\} $ such that $ g_k > f_k $, then $ j^* \ge 1 $.
        \item If $ h^* = |G^\prec(\alpha)| $, then $ |G^\prec_\dashv(\alpha)| = |G^\prec(\alpha)| $.
        \item $ |G^\prec_\dashv(\alpha)| \ge h^* $.
        \item If $ h^* > |G^\prec(\alpha)| $, $ |G^\prec(\alpha)| < f \le |Q| $ and there exists $ \beta \in I_{Q[f]} $ such that $ \beta \prec \alpha $, then $ Q[f] \in G_\dashv (\alpha) $.
        \item If $ h^* > |G^\prec(\alpha)| $, then $ |G^\prec_\dashv(\alpha)| = t^* $.
    \end{enumerate}

    Then, the conclusions will follow from point 5, point 6 and point 8.

    \begin{enumerate}
        \item Let us prove that, if $ i^* \ge 1 $, then $ Q[i^*] \in G_\dashv (\alpha) $. Since $ i^* \ge 1 $, we know that $ Q[i^*] \in G^* (\alpha) $, so $ Q[i^*] \in G_\dashv (\alpha) $.
        \item     Let us prove that, if $ j^* \ge 1 $, then $ Q[j^*] \in G_\dashv (\alpha) $. By the minimality of $ j^* $, we conclude that there exists $ 0 < k < \min\{r + 1,|\alpha|\} $ such that $ g_k > f_k $ and $ \mathtt{in}(Q[1, j^* - 1], s(\alpha, k)) < g_k $ (in particular, $ g_k \ge 1 $). Moreover, by the definition of $ j^* $ we have $ \mathtt{in}(Q[1, j^*], s(\alpha, k)) \ge g_k $. This implies that, for this $ k $, we have that $ j^* $ is the smallest integer $ 0 \le j \le |Q| $ such that $ \mathtt{in}(Q[1, j], s(\alpha, k)) \ge g_k $.
    
    Since $ g_k > f_k $, then there exist $ u \in G_\dashv(p(\alpha, |\alpha| - k)) $ and $ v \in Q $ such that $ (u, v, s(\alpha, k)) \in E $. Let $ u^* $ be the largest (w.r.t $ \le $) state $ u $ of $ G_\dashv(p(\alpha, |\alpha| - k)) $ such that $ (u, v, s(\alpha, k)) \in E $ for some $ v \in Q $, and let $ 1 \le x^* \le |Q| $ be the largest integer $ 1 \le x \le |Q| $ such that $ (u^*, Q[x], s(\alpha, k)) \in E $. From $ u^* \in G_\dashv(p(\alpha, |\alpha| - k)) $ we obtain that there exists $ \beta \in I_{u^*} $ such that $ p(\alpha, |\alpha| - k) \dashv \beta $, so from $ (u^*, Q[x^*], s(\alpha, k)) \in E $ we obtain $ \beta s(\alpha, k) \in I_{v^*} $ and $ \alpha = p(\alpha, |\alpha| - k) s(\alpha, k) \dashv \beta s(\alpha, k) $, so $ Q[x^*] \in G_\dashv (\alpha) $. The conclusion will follow if we prove that $ x^* = j^* $.

    Notice that for every $ (u, v, s(\alpha, k)) \in E $ such that $ v < Q[x^*] $ we have $ u \in G^\prec_\dashv(p(\alpha, |\alpha| - k)) $, because if we consider $ (u, v, s(\alpha, k)), (u^*, Q[x^*], s(\alpha, k)) \in E $, then by Axiom 4 we conclude $ u \le u^* $, so from $ u^* \in G_\dashv(p(\alpha, |\alpha| - k)) \subseteq G^\prec_\dashv(p(\alpha, |\alpha| - k)) $ we conclude $ u \in G^\prec_\dashv(p(\alpha, |\alpha| - k)) $. As a consequence, to prove that $ x^* = j^* $, we only have to prove that for every $ (u, v, s(\alpha, k)) \in E $ such that $ u \in G^\prec_\dashv(p(\alpha, |\alpha| - k)) $ we have $ v \le Q[x^*] $, because then from $ u^* \in G_\dashv(p(\alpha, |\alpha| - k)) $ and $ (u^*, Q[x^*], s(\alpha, k)) \in E $ we will conclude that $ x^* $ is the smallest integer $ 0 \le j \le |Q| $ such that $ \mathtt{in}(Q[1, j], s(\alpha, k)) \ge g_k $, and so $ x^* = j^* $.

    Fix $ (u, v, s(\alpha, k)) \in E $ such that $ u \in G^\prec_\dashv(p(\alpha, |\alpha| - k)) $. We must prove that $ v \le Q[x^*] $. Suppose for the sake of a contradiction that $ Q[x^*] < v $. If we consider $ (u^*, Q[x^*], s(\alpha, k)), (u, v, s(\alpha, k)) \in E $, then by Axiom 4 we conclude $ u^* \le u $. We distinguish two cases, and we show that both lead to a contradiction.
    \begin{itemize}
        \item Assume that $ u^* < u $. Then, from $ u^* \in G_\dashv(p(\alpha, |\alpha| - k)) $ and $ u \in G^\prec_\dashv(p(\alpha, |\alpha| - k)) $ we conclude $ u \in G_\dashv(p(\alpha, |\alpha| - k)) $. However, the edge $ (u, v, s(\alpha, k)) $ contradicts the maximality of $ u^* $.
        \item Assume that $ u^* = u $. Since $ (u, v, s(\alpha, k)) \in E $ and $ Q[x^*] < v $, then $ v $ contradicts the maximality of $ x^* $.
    \end{itemize}

    \item Let us prove that, if $ h^* = |G^\prec(\alpha)| $, then $ i^* = j^* = 0 $. It will suffice to show that, if $ (i^* \ge 1) \lor (j^* \ge 1) $, then $ h^* > |G^\prec(\alpha)| $. Assume that $ i^* \ge 1 $ (the case $ j^* \ge 1 $ is analogous). Then, by point 1 we have $ Q[i^*] \in G_\dashv (\alpha) $, so $ i^* > |G^\prec(\alpha)| $ and we conclude $ h^* \ge i^* > |G^\prec(\alpha)| $.

    \item Let us prove that, if there exists $ 0 < k < \min\{r + 1,|\alpha|\} $ such that $ g_k > f_k $, then $ j^* \ge 1 $. Notice that $ g_k \ge 1 $, so $\mathtt{in}(Q[1, 0], s(\alpha, k)) = 0 < g_k $ and the conclusion follows.

    \item Let us prove that, if $ h^* = |G^\prec(\alpha)| $, then $ |G^\prec_\dashv(\alpha)| = |G^\prec(\alpha)| $. We only have to prove that $ G_\dashv(\alpha) = \emptyset $. Suppose for the sake of a contradiction that $ G_\dashv(\alpha) \not = \emptyset $, and let $ u \in G_\dashv(\alpha) $. Since $ \alpha \not = \epsilon $, then there must exist $ 1 \le y \le |Q| $, $ u' \in Q $, $ \beta \in I_{u'}$, $ 0 < k \le r $ and $ \rho \in \Sigma^k $ such that (i) there exists an $ \epsilon $-walk from $ Q[y] $ to $ u $, (ii) $ (u', Q[y], \rho) \in E $ and (iii) $ \alpha \dashv \beta \rho $. We distinguish two cases, and we show that both lead to a contradiction.
    \begin{itemize}
        \item Assume that $ \alpha \dashv \rho $. Then, $ Q[y] \in G^*(\alpha) $, so $ i^* \ge y \ge 1 $. However, by point 3 we have $ i^* = 0 $, a contradiction.
        \item Assume that $ \lnot (\alpha \dashv \rho) $. Since $ \alpha \dashv \beta \rho $, then we have $ |\alpha| > k $, $ \rho = s(\alpha, k) $ and $ p(\alpha, |\alpha| - k) \dashv \beta $. In particular, we have $ 0 < k < \min\{r + 1,|\alpha|\} $, and from $ \beta \in I_{u'} $ we obtain $ u' \in G_\dashv(p(\alpha, |\alpha| - k)) $. From $ (u', Q[y], s(\alpha, k)) = (u', Q[y], \rho) \in E $ and $ u' \in G_\dashv(p(\alpha, |\alpha| - k)) $ we obtain $ g_k > f_k $, so from point 4 we obtain $ j^* \ge 1 $. However, by point 3 we have $ j^* = 0 $, a contradiction.
    \end{itemize}
    \item Let us prove that $ |G^\prec_\dashv(\alpha)| \ge h^* $. Since $ h^* = \max \{|G^\prec(\alpha)|, i^*, j^* \} $, we only need to prove that $ |G^\prec_\dashv(\alpha)| \ge |G^\prec(\alpha)| $, $ |G^\prec_\dashv(\alpha)| \ge i^* $ and $ |G^\prec_\dashv(\alpha)| \ge j^* $. The inequality $ |G^\prec_\dashv(\alpha)| \ge |G^\prec(\alpha)| $ is immediate, so let us prove that $ |G^\prec_\dashv(\alpha)| \ge i^* $ (the inequality $ |G^\prec_\dashv(\alpha)| \ge j^* $ can be proven analogously). If $ i^* = 0 $ we are done, so we can assume $ i^* \ge 1 $. By point 1 we have $ Q[i^*] \in G_\dashv(\alpha) $, so $ Q[i^*] \in G^\prec_\dashv(\alpha) $ and we conclude $ |G^\prec_\dashv(\alpha)| \ge i^* $. 

    \item Let us prove that, if $ h^* > |G^\prec(\alpha)| $, $ |G^\prec(\alpha)| < f \le |Q| $ and there exists $ \beta \in I_{Q[f]} $ such that $ \beta \prec \alpha $, then $ Q[f] \in G_\dashv (\alpha) $. Since $ h^* > |G^\prec(\alpha)| $, by point 6 we have $ |G^\prec_\dashv(\alpha)| > |G^\prec(\alpha)| $, so $ Q[|G^\prec(\alpha)| + 1] \in G_\dashv(\alpha) $. As a consequence, if $ f = |G^\prec(\alpha)| + 1 $ we are done. Now assume that $ f > |G^\prec(\alpha)| + 1 $. Since $ Q[|G^\prec(\alpha)| + 1] \in G_\dashv(\alpha) $, then there exists $ \gamma \in I_{Q[|G^\prec(\alpha)| + 1]} $ such that $ \alpha \dashv \gamma $, and in particular $ \alpha \preceq \gamma $, which implies $ \beta \prec \alpha \preceq \gamma $. From $ f > |G^\prec(\alpha)| + 1 $ and Axiom 1, we obtain $ Q[|G^\prec(\alpha)| + 1] \preceq_\mathcal{A} Q[f] $, so from $ \gamma \in I_{Q[|G^\prec(\alpha)| + 1]} $, $ \beta \in I_{Q[f]} $ and $ \beta \prec \gamma $ we obtain $ \{\beta, \gamma \} \subseteq I_{Q[|G^\prec(\alpha)| + 1]} \cap I_{Q[f]} $ and in particular $ \gamma \in I_{Q[f]} $, which implies $ Q[f] \in G_\dashv (\alpha) $.

    \item Let us prove that, if $ h^* > |G^\prec(\alpha)| $, then $ |G^\prec_\dashv(\alpha)| = t^* $.

    $ (\ge) $ Let us prove that for every $ 0 \le d \le t^* - h^* $ we have $ |G^\prec_\dashv(\alpha)| \ge h^* + d $. The conclusion will follow by choosing $ d = t^* - h^* $. We proceed by induction on $ d $. If $ d = 0 $, we have $ |G^\prec_\dashv(\alpha)| \ge h^* $ by point 6. Now, assume that $ 1 \le d \le t^* - h^* $ (which implies $ 1 \le h^* + d \le |Q| $ because $ 1 \le d \le h^* + d \le t^* \le |Q| $). To prove that $ |G^\prec_\dashv(\alpha)| \ge h^* + d $ it will suffice to prove that $ Q[h^* + d] \in G_\dashv (\alpha) $. Notice that $ h^* + 1 \le h^* + d \le t^* $, so by the minimality of $ t^* $ we have $ A_{\min}[h^* + d] < h^* + d $. This means that for some $ 1 \le g \le h^* + d - 1 $ there exists an $ \epsilon $-walk from $ Q[g] $ to $ Q[h^* + d] $. By the inductive hypothesis we know that $ |G^\prec_\dashv(\alpha)| \ge h^* + d - 1 $, so $ Q[g] \in G^\prec_\dashv(\alpha) $. Since $ G^\prec_\dashv(\alpha) $ is the disjoint union of $ G^\prec(\alpha) $ and $ G_\dashv(\alpha) $, we distinguish two cases.
    \begin{itemize}
        \item Assume that $ Q[g] \in G^\prec(\alpha) $. Pick any $ \beta \in I_{Q[g]} $. Since $ Q[g] \in G^\prec(\alpha) $, we have $ \beta \prec \alpha $. We know that there exists an $ \epsilon $-walk from $ Q[g] $ to $ Q[h^* + d] $, so $ \beta \in I_{Q[h^* + d]} $. Notice that $ h^* + d > h^* > |G^\prec(\alpha)| $, so by point 7 we conclude $ Q[h^* + d] \in G_\dashv (\alpha) $.
        \item Assume that $ Q[g] \in G_\dashv(\alpha) $. Then, there exists $ \beta \in I_{Q[g]} $ such that $ \alpha \dashv \beta $. We know that there exists an $ \epsilon $-walk from $ Q[g] $ to $ Q[h^* + d] $, so $ \beta \in I_{Q[h^* + d]} $, and we conclude $ Q[h^* + d] \in G_\dashv (\alpha) $.
    \end{itemize}

    $ (\le) $ If $ t^* = |Q| $ the conclusion is immediate, so in the following we assume $ t^* \le |Q| - 1 $. The conclusion will follow if we prove that $ Q[t^* + 1] \not \in G^\prec_\dashv(\alpha) $. Since $ G^\prec_\dashv(\alpha) $ is the disjoint union of $ G_\dashv(\alpha) $ and $ G^\prec(\alpha) $, we need to prove that $ Q[t^* + 1] \not \in G^\prec(\alpha) $ and $ Q[t^* + 1] \not \in G_\dashv(\alpha) $. From $ t^* + 1 > h^* > |G^\prec(\alpha)| $ we immediately conclude $ Q[t^* + 1] \not \in G^\prec(\alpha) $. Now, let us prove that $ Q[t^* + 1] \not \in G_\dashv(\alpha) $. Suppose for the sake of a contradiction that $ Q[t^* + 1] \in G_\dashv(\alpha) $. Since $ \alpha \not = \epsilon $, then there must exist $ 1 \le y \le |Q| $, $ u' \in Q $, $ \beta \in I_{u'}$, $ 0 < k \le r $ and $ \rho \in \Sigma^k $ such that (i) there exists an $ \epsilon $-walk from $ Q[y] $ to $ Q[t^* + 1] $, (ii) $ (u', Q[y], \rho) \in E $ and (iii) $ \alpha \dashv \beta \rho $. From the definition of $ t^* $ we obtain $ A_{\min}[t^* + 1] = t^* + 1 $, so we have $ t^* + 1 \le y $ and so $ h^* \le t^* < y $. We distinguish two cases, and we show that both lead to a contradiction.
        \begin{itemize}
            \item Assume that $ \alpha \dashv \rho $. This implies $ Q[y] \in G^*(\alpha) $, so $ i^* \ge y $. This yields a contradiction because $ i^* \le h^* < y $.
            \item Assume that $ \lnot (\alpha \dashv \rho) $. Since $ \alpha \dashv \beta \rho $, then we have $ |\alpha| > k $, $ \rho = s(\alpha, k) $ and $ p(\alpha, |\alpha| - k) \dashv \beta $. In particular, we have $ 0 < k < \min\{r + 1,|\alpha|\} $, and from $ \beta \in I_{u'}$ we obtain $ u' \in G_\dashv(p(\alpha, |\alpha| - k)) $. From $ (u', Q[y], s(\alpha, k)) = (u', Q[y], \rho) \in E $ and $ u' \in G_\dashv(p(\alpha, |\alpha| - k)) $ we obtain $ g_k > f_k $. Notice that for every $ (u, v, s(\alpha, k)) \in E $, if $ v < Q[y] $, then $ u \in G^\prec_\dashv(p(\alpha, |\alpha| - k)) $. Indeed, if we consider $ (u, v, s(\alpha, k)), (u', Q[y], s(\alpha, k)) \in E $, then by Axiom 4 we conclude $ u \le u' $, so from $ u' \in G_\dashv(p(\alpha, |\alpha| - k)) \subseteq G^\prec_\dashv(p(\alpha, |\alpha| - k)) $ we conclude $ u \in G^\prec_\dashv(p(\alpha, |\alpha| - k)) $. This implies $\mathtt{in}(Q[1, y - 1], s(\alpha, k)) < g_k $ because $ (u', Q[y], s(\alpha, k)) \in E $ and $ u' \in G^\prec_\dashv(p(\alpha, |\alpha| - k)) $. We conclude $ j^* \ge y $. This yields a contradiction because $ j^* \le h^* < y $. \hfill $ \qed $
        \end{itemize}
\end{enumerate}
\end{proof}

\end{document}